\documentclass[%
 reprint,
 amsmath,amssymb,
 aps,
 pre,
 floatfix,
]{revtex4-2}

\usepackage{graphicx}
\usepackage{dcolumn}
\usepackage{bm}
\usepackage{color}
\usepackage{tikz}
\usetikzlibrary{shapes,arrows}

\begin{document}

\preprint{APS/123-QED}




\title{Wave-Like Statistics from Classical Active Particles with Internal Degrees Of Freedom}


\author{Rahil N. Valani}\email{rahil.valani@physics.ox.ac.uk}
\affiliation{Rudolf Peierls Centre for Theoretical Physics, Parks Road,
University of Oxford, OX1 3PU, United Kingdom}

\date{\today}

\begin{abstract}

Wave-like spatial statistics in walking-droplet systems are often associated with wave-mediated interactions and wave-memory effects. Here we explore how similar statistical structure can arise from the low-dimensional nonlinear dynamics of an inertial active particle with internal degrees of freedom. In this framework, steady propulsion corresponds to internal-state fixed points whose spiral or transiently chaotic relaxation organizes oscillatory ensemble densities. Local perturbations then generate wave-like statistics in both open and closed geometries, suggesting that wave-like ensemble behavior may emerge more generally from internal-state attractor dynamics in inertial active matter.

\end{abstract}

\maketitle

\section{Introduction}

Active particles are intrinsically nonequilibrium units that convert energy into persistent motion. Collections of such particles display a rich spectrum of emergent behaviors, including flocking and swarming~\citep{VICSEK201271}, motility-induced phase separation~\citep{doi:10.1146/annurev-conmatphys-031214-014710}, and active turbulence~\citep{Doostmohammadi2018}. While microscopic realizations are often well described by overdamped dynamics, inertia becomes significant for larger particles leading to underdamped active behavior~\citep{lowen2020}. The hydrodynamic active systems of walking~\citep{Couder2005WalkingDroplets} and superwalking droplets~\citep{superwalker} provide a striking physical realization of such inertial active particles. In these systems, an oil droplet bouncing synchronously on a vertically vibrated bath becomes self-propelled horizontally through interaction with a self-generated wave field~\citep{molacek_bush_2013,Molacek2013DropsTheory,Protiere2006}. The resulting motion is inherently non-Markovian, as propulsion depends on the cumulative wave field generated along the droplet’s past trajectory~\citep{Oza2013}. Owing to this wave–particle coupling, walking droplets have been shown to exhibit a range of hydrodynamic quantum analogs, including the emergence of wave-like ensemble statistics~\citep{Bush2015,Bush2020review,Bush2024,Bush2024APL}.

Two prominent manifestations of wave-like statistics in this system are: (i) the hydrodynamic analog of Friedel oscillations~\citep{Saenz2020}, where spatially decaying density modulations arise near localized defects in open geometries, and (ii) the emergence of coherent wave-like positional statistics in confined droplet geometries such as circular corrals ~\citep{PhysRevE.88.011001,Saenz2017}. In these experiments, wave-like ensemble structure emerges from complex wave–particle dynamics involving wave–defect or wave–boundary interactions. At the same time, previous studies have established that oscillatory relaxation of the droplet speed is itself an intrinsic feature of pilot-wave dynamics at sufficiently high memory~\citep{Durey2020lorenz,Bacot2019,Blitstein2025arxiv,ValaniUnsteady}. This raises a broader question: \emph{are wave-like ensemble statistics a special consequence of wave-mediated interactions and wave-memory effects in the walking droplet system, or do they reflect a more general dynamical mechanism applicable to a wider class of inertial active particles?}

To investigate this question, we develop an internal-state description of an inertial active particle whose self-propulsion is governed by a low-dimensional Lorenz-type dynamical system. This framework is motivated by the exact reduction of the walking-droplet integro-differential trajectory equation~\citep{Oza2013} to a local Lorenz-type system~\citep{Durey2020lorenz,ValaniUnsteady,Valanilorenz2022,Lorenz1963}, which has proven effective in probing the dynamical origins of a range of hydrodynamic quantum analogs~\citep{VALANI2024115253,MegastableLopezValani2025,Perks2023,ZuValani2025,Valani2022ANM}. Within this formulation, steady propulsion corresponds to fixed points of the internal-state dynamics, whose stability governs the particle's response to localized perturbations. We show that stable spiral equilibria and transient chaotic attractors organize oscillatory ensemble density modulations in both open and confined geometries. By demonstrating that the same dynamical mechanism also arises in a different active-particle model not derived from walking-droplet dynamics, we suggest that wave-like ensemble statistics may be a broader consequence of low-dimensional internal-state attractors governing inertial active motion.


\begin{figure}
\centering
\includegraphics[width=0.85\columnwidth]{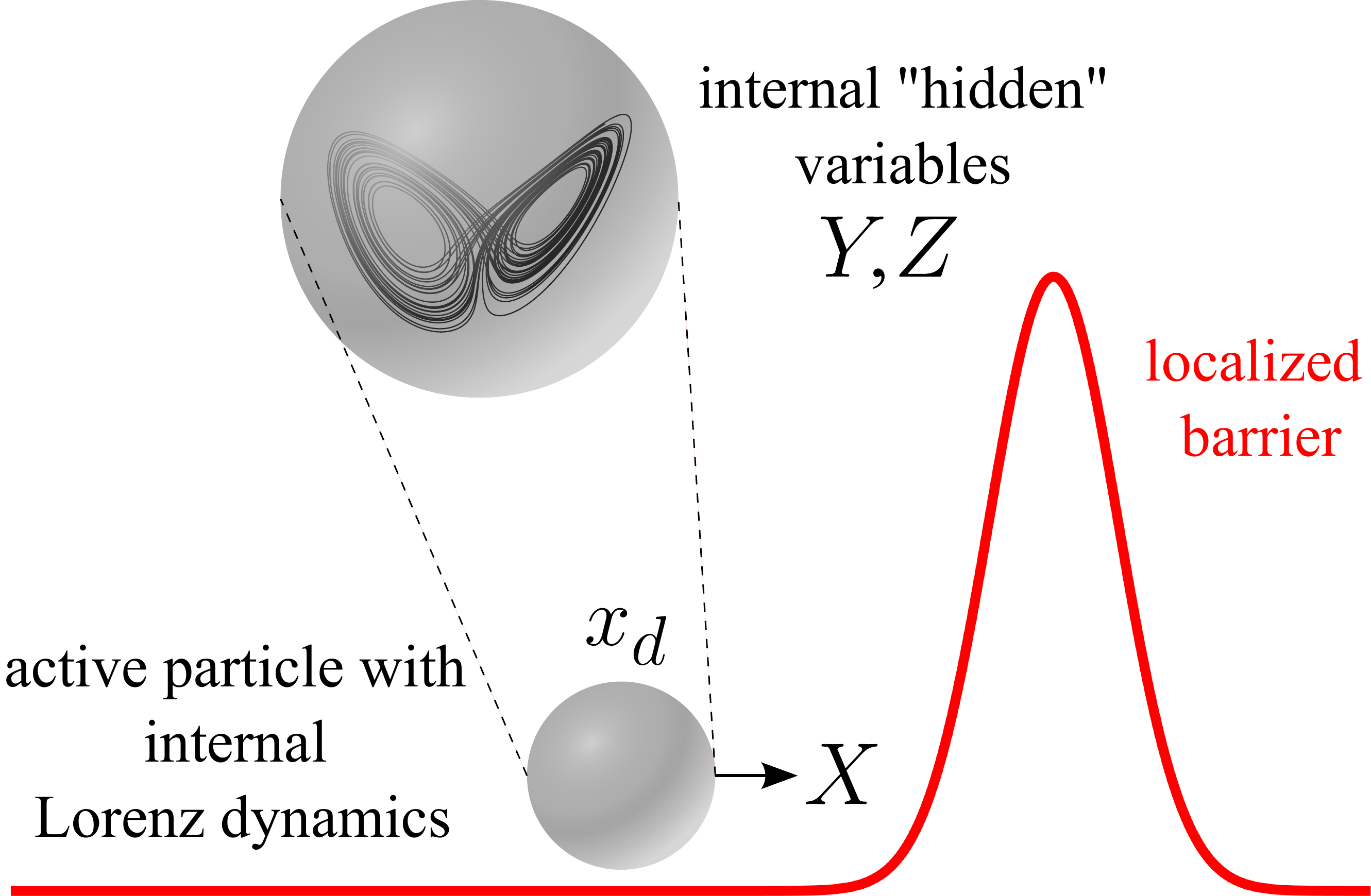}
\caption{Inertial active particle with internal Lorenz dynamics. An inertial active particle at position $x_d$ moves in one dimension with velocity $X$ and interacts with a localized external potential barrier. Its self-propulsion is governed by two internal-state (“hidden”) variables $Y$ and $Z$ that obey a Lorenz-type dynamical system.}
\label{fig:schematic}
\end{figure}

\begin{figure*}
\centering
\includegraphics[width=1.5\columnwidth]{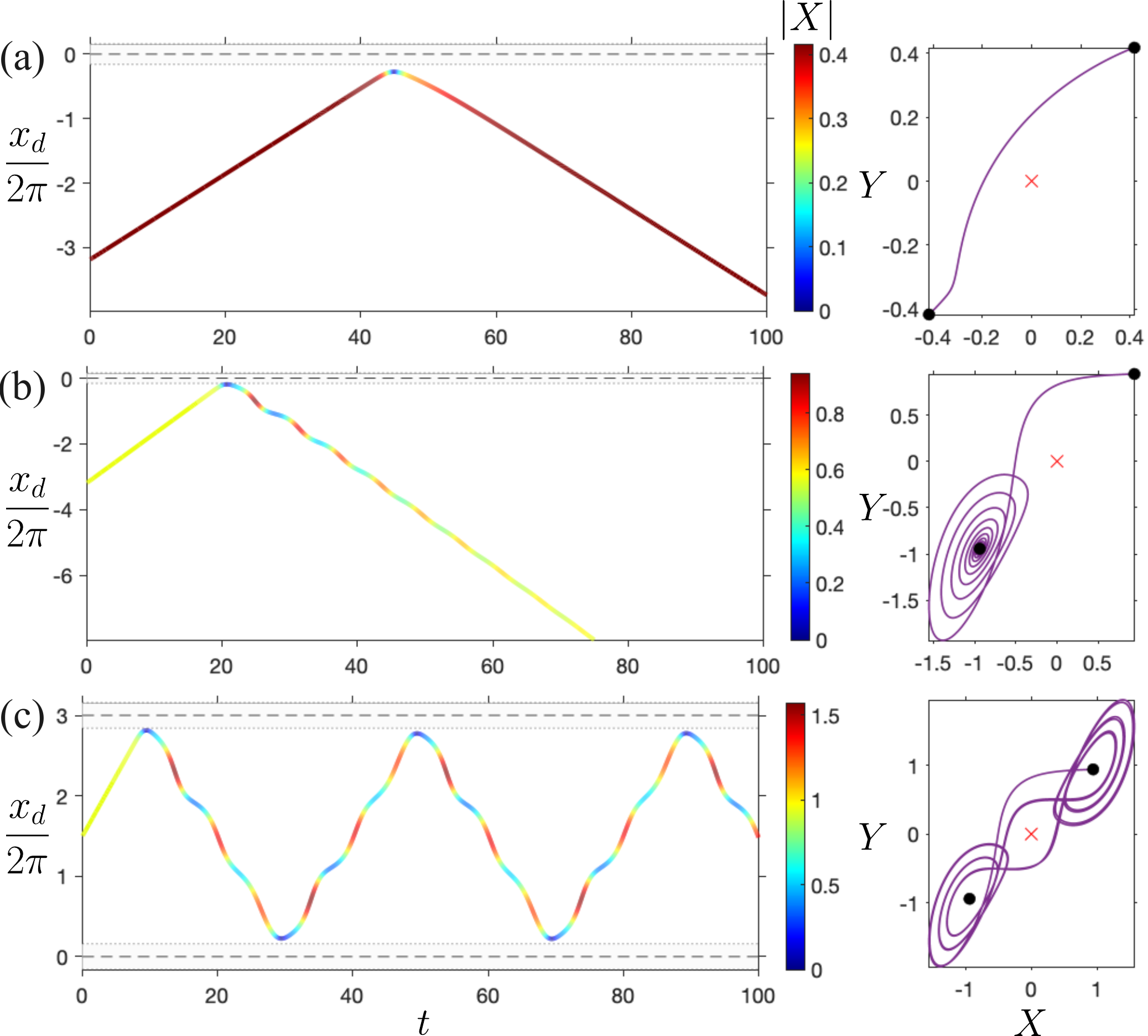}
\caption{Active particle response to perturbations in open and closed geometries. 
Left: space–time trajectories $x_d(t)$ (color indicates speed $|X|$), right: corresponding $(X,Y)$ phase-plane projections. 
(a,b) Open geometry: interaction with a single Gaussian barrier at $x_b=0$ (dashed line at $x_b$, dotted lines at $x_b\pm W$). 
(a) $\tau=1.1$, the steady-propulsion equilibrium is a stable node in internal state-space and perturbations relax monotonically. 
(b) $\tau=3$, the equilibrium is a stable spiral, producing underdamped oscillatory relaxation after scattering. 
(c) Closed geometry: two Gaussian barriers at $x_b=0$ and $x_b=6\pi$ (“particle-in-a-box”), yielding persistent speed oscillations for $\tau=3$. 
Black dots denote steady-propulsion equilibria $(X^*,Y^*)=(\pm\sqrt{R-1/\tau^2},\pm\sqrt{R-1/\tau^2})$; the red cross marks the stationary state saddle $(X^*,Y^*)=(0,0)$. 
Other parameters are $R=1$, $W=1$, $H=5$.}
\label{fig:single}
\end{figure*}

\section{Internal-state model of an inertial active particle}

To investigate the dynamical origin of wave-like ensemble statistics, we consider an inertial active particle whose self-propulsion is governed by a small number of internal degrees of freedom. The particle evolves under the combined action of inertia, drag, an external potential, and an intrinsic propulsion force generated by its internal state. Our goal is to identify the generic dynamical features of such internal-state systems that organize wave-like ensemble statistics. As we show below, the reduced equations describing walking droplets provide one particular realization of this broader class of active particles.

A broad class of inertial active particles can be described by coupling the particle motion to a small number of internal variables that evolve according to nonlinear dissipative dynamics~\citep{Valaniattractormatter2023}. In such systems, the internal state continuously regulates the propulsion force acting on the particle, while external perturbations modify the particle trajectory through localized forcing. A simple realization of this idea is obtained when the internal-state dynamics are governed by a Lorenz-type system~\citep{Lorenz1963,ZuValani2025}, leading to

\begin{equation}
\begin{aligned}
\dot{x}_d &= X, \\
\dot{X} &= Y - X + H(x_d-x_b)e^{-(x_d-x_b)^2/W^2}, \\
\dot{Y} &= -\frac{1}{\tau}Y + XZ, \\
\dot{Z} &= R - XY - \frac{1}{\tau}Z.
\end{aligned}
\label{eq:lorenz}
\end{equation}

The first two equations describe the particle motion: $\dot{x}_d=X$ is the kinematic relation between position and velocity, while the $\dot{X}$ equation represents a force balance in which the particle experiences a linear drag force $-X$, an intrinsic propulsion force generated by the internal-state variable $Y$, and a localized external force due to the Gaussian potential. The remaining two equations describe the nonlinear evolution of the internal-state variables $(Y,Z)$, which regulate the self-propulsion dynamics through their coupling to the particle velocity. The external potential is taken to be a localized Gaussian barrier
\[
V(x)=V_0e^{-(x-x_b)^2/W^2},
\]
whose corresponding force
\[
F_{\rm ext}
=
H(x_d-x_b)e^{-(x_d-x_b)^2/W^2}
\]
acts as a localized perturbation to the internal-state dynamics, where $H=2V_0/W^2$ and $W$ characterize the amplitude and width of the Gaussian bump, respectively. Figure~\ref{fig:schematic} illustrates this setup.

The active-particle model considered here is motivated by the reduced dynamics of walking droplets (see Appendix~\ref{sec:1}). In particular, the Lorenz-type system in Eq.~\eqref{eq:lorenz} arises through an exact reduction of the one-dimensional stroboscopic pilot-wave equation when a sinusoidal interaction kernel is adopted~\citep{Durey2020lorenz,ValaniUnsteady,Valanilorenz2022}. Within this derivation, the internal-state variables have a direct physical interpretation: $Y$ corresponds to the wave-memory force acting on the droplet and $Z$ to the local wave height. In the present work, however, we adopt the broader active-particle interpretation, treating $(Y,Z)$ as generic internal degrees of freedom governing self-propulsion. This perspective allows the dynamical mechanism identified below to be viewed independently of its particular walking-droplet realization.

In the absence of external perturbations ($|x_d-x_b|\gg W$), Eq.~\eqref{eq:lorenz} reduces to the rescaled Lorenz system~\cite{Lorenz1963,Valani2024}. Steady propulsion corresponds to fixed points of the internal-state dynamics and the Lorenz phase-space structure organizes the particle’s response to perturbations. The localized potential acts as a transient forcing of the $\dot{X}$ equation, displacing the system in phase space and initiating internal-state relaxation. 

The unperturbed system (no Gaussian barrier) admits a stationary non-walking equilibrium at $(X,Y,Z)=(0,0,\tau R)$ and, for $\tau>1/\sqrt{R}$, a symmetric pair of steady-propulsion equilibria
\[
(X^*,Y^*,Z^*)=\left(\pm\sqrt{R-\tfrac{1}{\tau^2}},\ \pm\sqrt{R-\tfrac{1}{\tau^2}},\ \tfrac{1}{\tau}\right),
\]
corresponding to steady self-propulsion~\citep{Valani2024,ZuValani2025}. Linearization about the steady-propulsion state yields eigenvalues that are either real (stable node) or complex with negative real part (stable spiral), depending on $(R,\tau)$. In the spiral regime, small perturbations generate underdamped oscillatory relaxation in $(X,Y,Z)$, which manifests physically as transient velocity oscillations.


\begin{figure*}
\centering
\includegraphics[width=1.2\columnwidth]{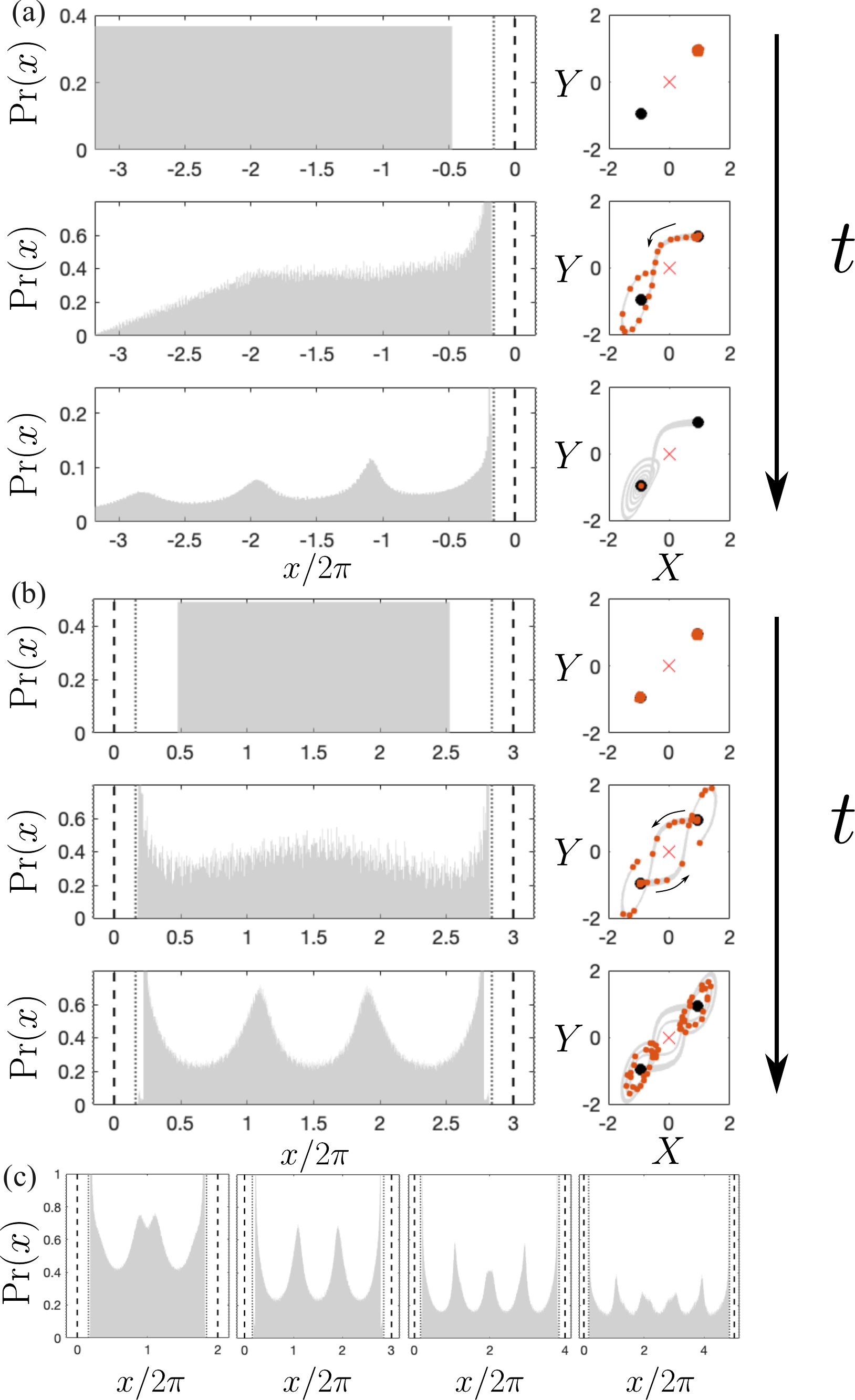}
\caption{Emergence of wave-like ensemble statistics from internal Lorenz dynamics in open and closed geometries. 
Left panels show the evolving position probability density $\mathrm{Pr}(x)$ constructed from an ensemble of 1000 trajectories; right panels show the corresponding evolution in the $(X,Y)$ phase plane. $\Pr(x)$ denotes cumulative ensemble probability densities. 
(a) Open geometry (single barrier). From top to bottom: $t=0$, $t=8.2$, and $t=200$. All particles are initialized with velocities directed toward the barrier. 
(b) Closed geometry (two barriers). From top to bottom: $t=0$, $t=7.9$, and $t=500$. Particles are initialized with half of the velocities directed left and half right. 
(c) Final probability distributions in the closed geometry for different box lengths (left to right): $4\pi$, $6\pi$, $8\pi$, and $10\pi$. 
Black dots denote steady-propulsion equilibria and the red cross denotes the saddle at the origin. Other parameters are same as Fig.~\ref{fig:single}.
}
\label{fig:ensemble}
\end{figure*}

\begin{figure*}
\centering
\includegraphics[width=1.8\columnwidth]{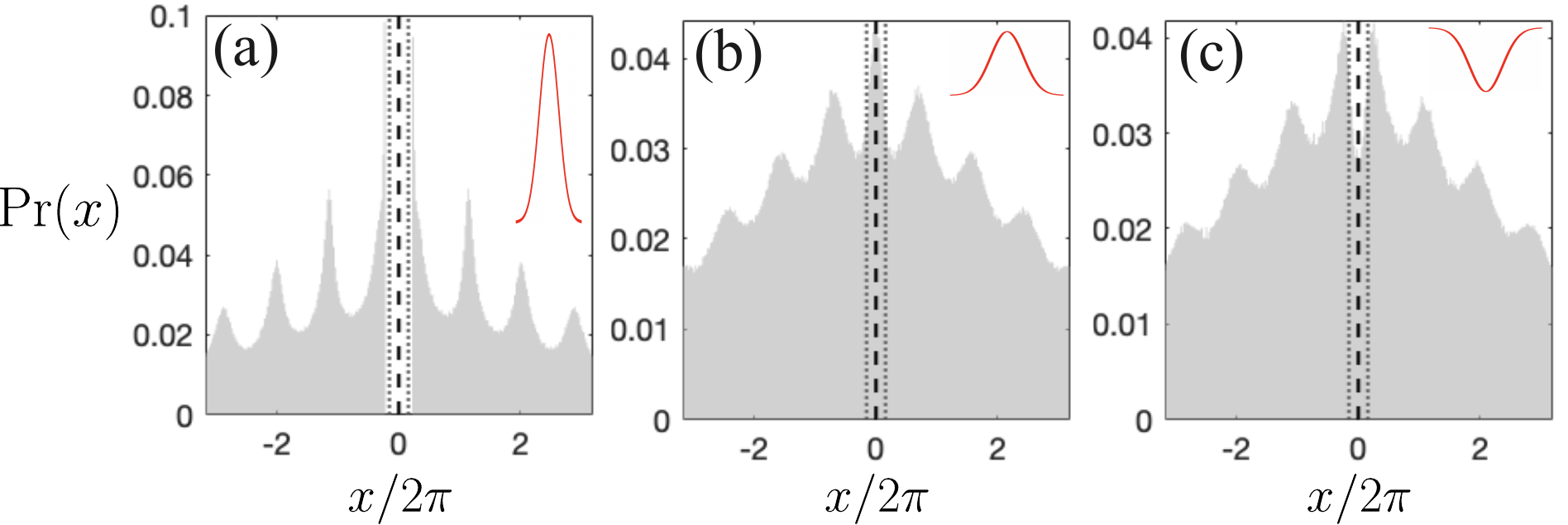}
\caption{Robustness of wave-like statistics to the barrier geometry. One-dimensional results obtained from the minimal Lorenz-like model for different bump geometries (dashed line at $x_b$, dotted lines at $x_b\pm W$): (a) a steep reflective barrier with $H=10$, (b) a shallow transmissive barrier with $H=1$, and (c) a shallow well with $H=-1$. In all one-dimensional cases, particles are initialized on either side of the defect and directed toward it. In these one-dimensional cases, the plotted distributions $\Pr(x)$ are cumulative probability densities at $t=200$ constructed from the ensemble history as time evolves. Parameters are $R=1$, $\tau=3$, $W=1$, $x_b=0$.}
\label{fig:robustness barrier}
\end{figure*}

\section{Results}

We numerically integrate Eq.~\eqref{eq:lorenz} using \texttt{ode45} in MATLAB and first examine the response of a single active particle to localized perturbations. Figure~\ref{fig:single} illustrates representative trajectories in both open and confined geometries. In the open configuration, a particle incident on a single Gaussian barrier is displaced from its steady-propulsion equilibrium. When the equilibrium is a stable node ($\tau=1.1$, Fig.~\ref{fig:single}(a)), the internal-state perturbation relaxes monotonically and no oscillatory motion develops. In contrast, when the equilibrium is a stable spiral ($\tau=3$, Fig.~\ref{fig:single}(b)), the barrier induces a transient phase-space excursion followed by underdamped spiral relaxation, generating oscillations in the velocity $X(t)$ that manifest as wave-like spatial modulation of particle density along its trajectory. In the confined geometry (Fig.~\ref{fig:single}(c)), where the particle motion is confined between two Gaussian barriers (“particle-in-a-box”), the same spiral attractor structure produces sustained speed oscillations and persistent wave-like particle density modulation within the box. Thus, the emergence of oscillations in spatial density is governed by the stability type of the internal-state equilibrium.

The oscillatory particle trajectories in Fig.~\ref{fig:single}(b,c) exhibit a characteristic spatial period that is close to the wavelength, $2\pi$, of the particle-generated waves in the corresponding walking-droplet interpretation~(see Appendix~\ref{sec:1}). To gain further insight into this length scale, we follow \citet{Durey2020} and compare it with the prediction from linear stability analysis of the free steadily propagating state. Linearization about the steady-propulsion equilibrium yields an oscillation frequency
\[
\omega=\sqrt{R+\frac{1}{\tau}},
\]
for small perturbations~(see the Appendix of Ref.~\citep{Valani2024}). Assuming that the particle propagates with approximately the steady speed $X^*=\sqrt{R-1/\tau^2}$ during this relaxation, the corresponding spatial oscillation wavelength is estimated as
$\Lambda \approx \frac{2\pi X^*}{\omega}$,
or equivalently,
\[
\frac{\Lambda}{2\pi}
\approx
\sqrt{\frac{R-\frac{1}{\tau^{2}}}{R+\frac{1}{\tau}}}.
\]
For the parameters used in Fig.~\ref{fig:single}(b), $R=1$ and $\tau=3$, this estimate gives $\Lambda/(2\pi)\approx0.82$.

Having established the single-particle mechanism, we now examine its consequences for ensemble statistics. Figure~\ref{fig:ensemble} illustrates how wave-like ensemble statistics emerge from the underlying internal Lorenz dynamics in both open and confined geometries. The left panels show the cumulative position probability density $\Pr(x)$ constructed from an ensemble of $1000$ trajectories, while the right panels display the corresponding evolution of the ensemble in the $(X,Y)$ phase plane. In the open geometry (Fig.~\ref{fig:ensemble}(a)), particles are initialized at different distances from a single barrier with velocities directed toward it. As the ensemble interacts with the barrier ($t=8.2$), trajectories are displaced from positive velocity equilibrium and spread along the spiral structure of the attractor. At long times ($t=200$), the ensemble relaxes toward the negative velocity stable equilibria through underdamped phase-space motion, and the resulting velocity oscillations organize spatially oscillatory modulations of $\Pr(x)$ characteristic of an open geometry in the corresponding walking-droplet analog of Friedel oscillations~\citep{Saenz2020}.

In the confined configuration (Fig.~\ref{fig:ensemble}(b)), where particles evolve between two barriers with initial velocities distributed left and right, the same dynamical mechanism produces wave-like ensemble statistics. 
Figure~\ref{fig:ensemble}(c) shows the final probability densities for different box lengths, demonstrating that confinement can organize structured wave-like ensemble statistics at characteristic length scales, while the underlying dynamical mechanism remains unchanged. We note that qualitatively similar wave-like statistics have been reported for walking droplets confined by a one-dimensional harmonic potential~\citep{durey2018}; here we emphasize a complementary dynamical interpretation based on internal-state attractor dynamics. We emphasize that these results are not intended to identify the mechanism responsible for the wave-like statistics observed in the original walking-droplet corral experiments~\citep{PhysRevE.88.011001}. Rather, they demonstrate that local perturbations of an active particle's internal-state dynamics provide a plausible alternative route by which wave-like ensemble statistics can emerge in confined geometries.

\begin{figure*}
\centering
\includegraphics[width=1.75\columnwidth]{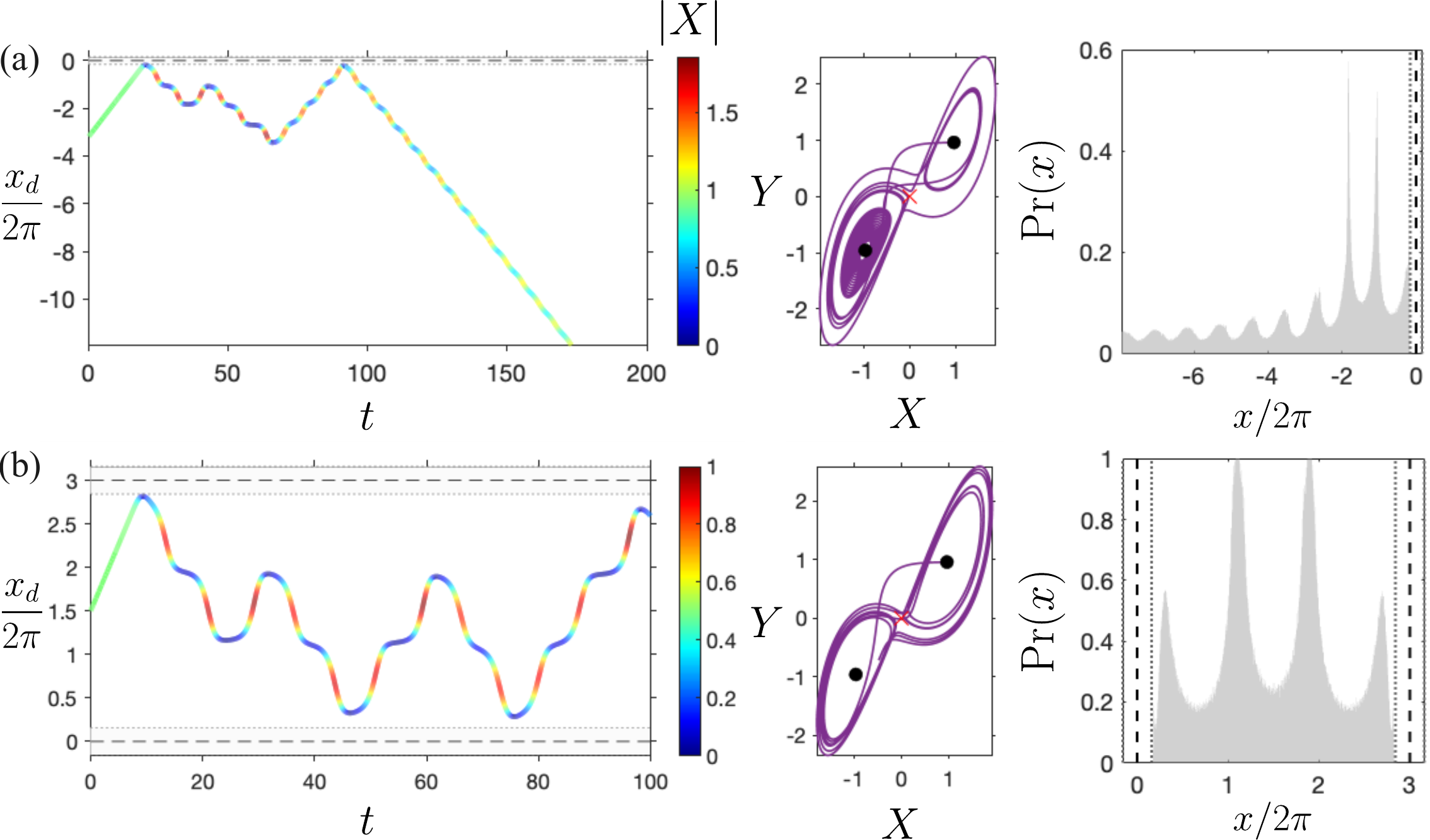}
\caption{Transient chaos from perturbations generates wave-like statistics. 
(a) Open geometry; (b) closed geometry with box-width $6\pi$. 
Left panels show space–time trajectories $x_d(t)$ (color indicates $|X|$), middle panels show the corresponding $(X,Y)$ phase-plane dynamics, and right panels show the final position probability density $\Pr(x)$ at $t=200$. 
Following barrier-induced perturbation, the dynamics depart from steady propulsion and evolve through transient chaotic motion before settling, producing coherent wave-like spatial probability distributions. Parameters: $\tau=3.6$; all other parameters as in Fig.~\ref{fig:single}.}
\label{fig:chaotic}
\end{figure*}

We next examine the robustness of the proposed mechanism to the geometry and sign of the localized perturbation. Figure~\ref{fig:robustness barrier} shows the long-time ensemble statistics for three representative potential landscapes: a steep repulsive barrier, a shallow transmissive barrier, and an attractive well, while keeping the internal-state dynamics unchanged. Although the detailed form of the probability density depends on the strength and sign of the perturbation, oscillatory ensemble statistics with a common characteristic wavelength persist in all cases. These results demonstrate that the emergence of wave-like statistics is remarkably insensitive to the detailed form of the localized forcing and is instead governed by the underlying internal-state dynamics. More generally, they indicate that wave-like ensemble structure is organized by the nature of the attractor of the internal-state dynamics rather than by the specific geometry of the external perturbation.



Transient chaotic dynamics provide a second route to the emergence of wave-like ensemble statistics. For $\tau=3.6$, localized perturbations displace the system away from steady propulsion into a regime of transient chaos before eventual relaxation to the steady-propulsion state. In both the open geometry (Fig.~\ref{fig:chaotic}(a)) and the confined configuration (Fig.~\ref{fig:chaotic}(b)), the internal-state trajectory undergoes chaotic excursions across the Lorenz attractor, generating irregular yet correlated fluctuations in the particle velocity. These fluctuations organize structured spatial probability distributions that retain a clear wave-like character despite the absence of simple periodic relaxation. Thus, coherent ensemble statistics need not arise solely from underdamped spiral relaxation near stable equilibria; they can emerge more generally from the geometry of the underlying attractor, including regimes of low-dimensional transient chaos.

\begin{figure*}
\centering
\includegraphics[width=2\columnwidth]{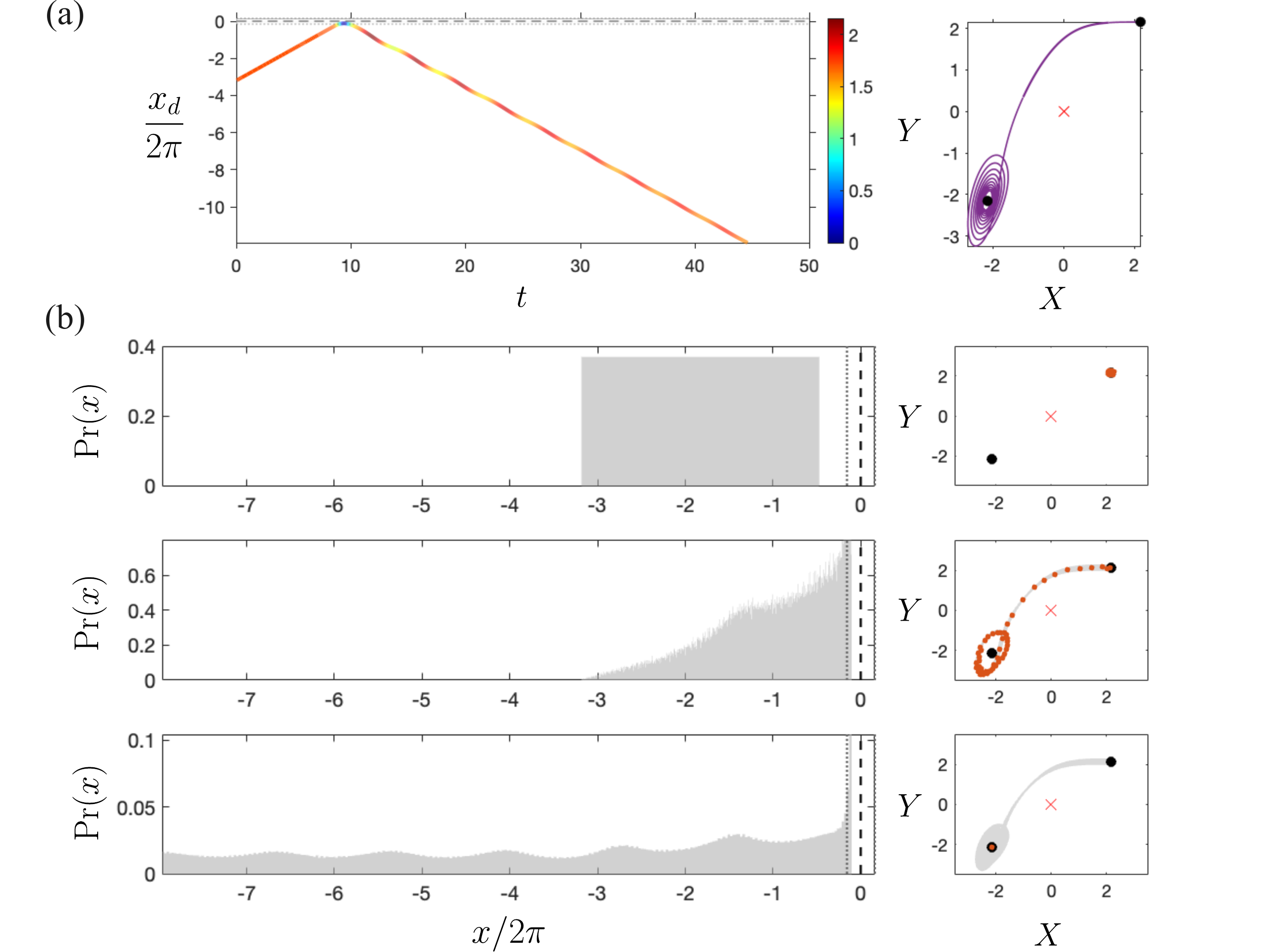}
\caption{Robustness of speed oscillations and wave-like statistics to internal-state dynamics. Considering an alternate internal state dynamics as given by Eq.~\eqref{eq:lorenz alt}, which is not derived from wave-particle coupling of walking droplets, still exhibits wave-like statistics. (a) Left: space–time trajectories $x_d(t)$ (color indicates speed $|X|$), right: corresponding $(X,Y)$ phase-plane projections. (b) Left panels show the evolving position probability density $\mathrm{Pr}(x)$ constructed from an ensemble of 1000 trajectories; right panels show the corresponding evolution in the $(X,Y)$ phase plane. $\Pr(x)$ denotes cumulative ensemble probability densities. From top to bottom: $t=0$, $t=8.2$, and $t=200$. All particles are initialized with velocities directed toward the barrier. Black dots denote steady-propulsion equilibria $(X^*,Y^*)=(\pm(\gamma/\beta)\sqrt{\alpha-1},\pm(\gamma/\beta)\sqrt{\alpha-1})$; the red cross marks the stationary state saddle $(X^*,Y^*)=(0,0)$. 
Parameters: $R=1$, $W=1$, $H=10$, $\alpha=4.1$, $\beta=1$ and $\gamma=1.5$.}
\label{fig:robustness internal-state}
\end{figure*}

To assess whether the proposed mechanism is specific to the walking-droplet reduction in Eq.~\eqref{eq:lorenz}, we consider an alternative active-particle model whose internal-state dynamics are not derived from wave--particle coupling but retain nonlinear feedback between the particle velocity and internal variables:
\begin{equation}
\begin{aligned}
\dot{x}_d &= X, \\
\dot{X} &= Y - X + H(x_d-x_b)e^{-(x_d-x_b)^2/W^2}, \\
\dot{Y} &= -Y - XZ + \alpha X, \\
\dot{Z} &= \beta X^2 - \gamma Z.
\end{aligned}
\label{eq:lorenz alt}
\end{equation}
Unlike Eq.~\eqref{eq:lorenz}, this system is introduced phenomenologically and is not obtained through an exact reduction of the walking-droplet equations. Nevertheless, it possesses steadily propagating states whose stability can be tuned independently through the parameters $(\alpha,\beta,\gamma)$.

Figure~\ref{fig:robustness internal-state} demonstrates that the same dynamical mechanism persists in this more general setting. A localized perturbation displaces the particle from its steady-propulsion equilibrium, producing underdamped spiral relaxation in the internal-state variables and the associated oscillations in the particle velocity (Fig.~\ref{fig:robustness internal-state}(a)). The corresponding ensemble evolution (Fig.~\ref{fig:robustness internal-state}(b)) closely mirrors that of the walking-droplet reduction: trajectories spread along the spiral attractor following the perturbation and subsequently relax back toward the steady-propulsion state, organizing oscillatory spatial probability densities. The emergence of wave-like ensemble statistics therefore does not rely on the specific form of the internal-state equations inherited from the walking-droplet model, but rather on the existence of stable spiral attractors governing the self-propulsion dynamics.

Linear stability analysis of the steady-propulsion equilibrium (see Appendix~\ref{sec:altstability}) shows that small perturbations oscillate with angular frequency $\omega$, where $\omega$ is determined by the complex eigenvalues of the internal-state dynamics. Assuming that the particle propagates at approximately the steady speed $X^*$ during this relaxation, the corresponding spatial oscillation wavelength is estimated as $\Lambda\approx2\pi X^*/\omega$. For the parameters used in Fig.~\ref{fig:robustness internal-state}, this estimate gives $\Lambda/(2\pi)\approx1.31$, consistent with the spacing of the oscillatory probability modulations observed numerically.

Finally, the attractor-driven mechanism identified here is not restricted to the one-dimensional active-particle model. Appendix~\ref{sec:2D} demonstrates that qualitatively similar wave-like ensemble statistics are also obtained in the full two-dimensional stroboscopic walking-droplet model, indicating that the phenomenon is not an artifact of the reduced description.

\section{Discussion and Conclusions}

We have identified a dynamical mechanism by which wave-like ensemble statistics emerge from the low-dimensional attractor structure of an inertial active particle with internal degrees of freedom. Within this framework, steady propulsion corresponds to fixed points of the internal-state dynamics, whose stability governs the particle's response to localized perturbations. Stable spiral equilibria generate underdamped phase-space relaxation that organizes oscillatory ensemble density modulations, while transient chaotic attractors provide an alternative route to coherent wave-like structure. In both open and confined geometries, the emergence of wave-like ensemble statistics is therefore governed by the nature of the underlying internal-state attractor rather than by the detailed form of the external perturbation.

Walking droplets provide a natural physical realization of this mechanism. Previous studies have shown that oscillatory speed dynamics are an intrinsic feature of pilot-wave systems at sufficiently high memory and that such oscillations underlie several hydrodynamic quantum analogs~\citep{Durey2020,Bacot2019,Blitstein2025arxiv}. The present work offers a complementary dynamical interpretation by relating these oscillations directly to the nature of the underlying internal-state attractor. This viewpoint also suggests an alternative experimental route for probing the mechanism using localized external forcing, for example through magnetic confinement~\citep{Perrard2014a}, in regimes where freely propagating walkers exhibit intrinsic speed oscillations~\citep{Bacot2019}. Such experiments could isolate the role of internal-state dynamics from the detailed wave--topography interactions present in conventional walking-droplet experiments.

More broadly, this work places walking droplets within the wider framework of inertial active matter. The exact reduction of the stroboscopic equations reveals that the walker can be viewed as an active particle whose self-propulsion is regulated by low-dimensional internal dynamics. Together with the alternate internal-state model presented in this work, these results suggest that wave-like ensemble statistics need not be regarded as exclusive to pilot-wave systems, but may instead represent a more general consequence of internal-state attractors governing active motion. We hope that this perspective motivates further investigation of wave-like statistical phenomena, and more generally hydrodynamic quantum analogs, in broader classes of inertial active particles with internal-state dynamics.

\section*{Acknowledgments} 
R.V. acknowledges the support of the Leverhulme Trust [Grant No. LIP-2020-014]. 

\bibliography{Wavelike_active_particle}

\appendix 

\begin{figure}[t]
\centering
\includegraphics[width=\columnwidth]{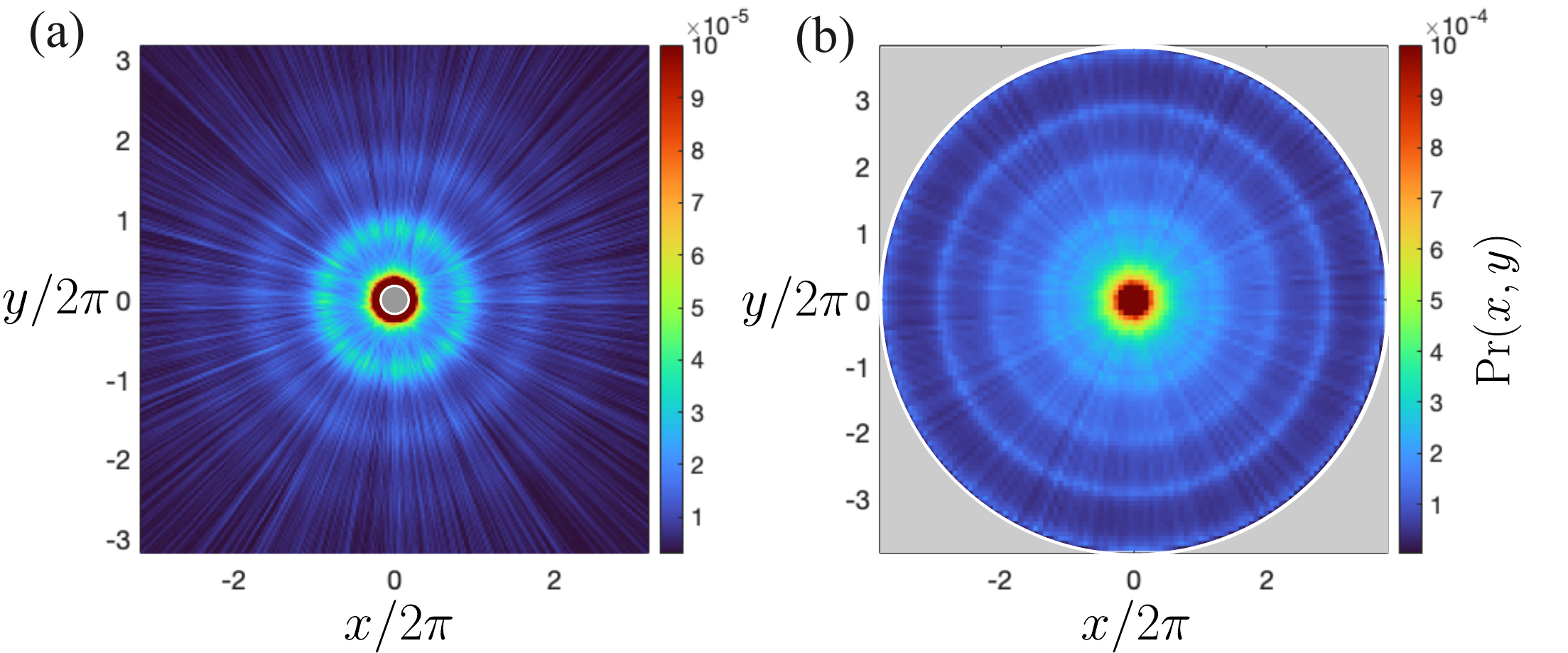}
\caption{Two-dimensional wave-like ensemble statistics. 
(a) Open geometry: particles initialized far from a localized bump at the origin evolve toward the bump radially inwards with small orientational noise, producing concentric Friedel-like oscillations in the probability density $\Pr(x,y)$ of outgoing trajectories for $\tau=3$. 
(b) Confined geometry: particles uniformly initialized inside a circular corral of radius $8\pi$ and directed radially outward with small orientational noise, generate wave-like spatial structure in $\Pr(x,y)$ for $\tau=4$. 
Color maps show $\Pr(x,y)$ constructed from $10^4$ trajectories. 
Parameters: $R=1$, $W=1$, $H=5$, $L=2\pi$.}
\label{fig:2D results}
\end{figure}

\section{Walking-droplet model reduction}\label{sec:1}

The active-particle model developed in the main text is motivated by the reduced dynamics of walking droplets. Here we summarize the derivation of the corresponding internal-state equations from the one- and two-dimensional stroboscopic walking-droplet model. 

We consider a droplet bouncing periodically on a vertically vibrated bath of the same liquid while moving horizontally. Since the time scale of vertical bouncing is much shorter than that of horizontal motion, the dynamics can be described using a stroboscopic approximation that averages over the vertical motion and yields a continuum model for horizontal walking~\citep{Oza2013}. In this framework, the droplet is treated as a point particle at horizontal position $\mathbf{x}_d(t)$, generating axisymmetric standing waves that decay exponentially in time. The horizontal dynamics are governed by
\begin{equation}
m\ddot{\mathbf{x}}_d + D\dot{\mathbf{x}}_d = -mg\nabla h(\mathbf{x}_d,t),
\label{eq:dim1}
\end{equation}
where $m$ is the droplet mass, $D$ is an effective drag coefficient, and $h(\mathbf{x},t)$ is the wave field generated by the particle. The wave field is constructed from the superposition of individual waves $W(|\mathbf{x}|)$ generated along the particle’s past trajectory,
\begin{equation}
h(\mathbf{x},t)=\frac{A}{T_F}\int_{-\infty}^{t} 
W\!\left(k_F|\mathbf{x}-\mathbf{x}_d(s)|\right)
e^{-(t-s)/(T_F\mathrm{Me})}\,\mathrm{d}s,
\label{eq:dim2}
\end{equation}
where $k_F$ is the Faraday wavenumber, $A$ is the wave amplitude, $\mathrm{Me}$ is the memory parameter, and $T_F$ is the Faraday period. Substituting Eq.~\eqref{eq:dim2} into Eq.~\eqref{eq:dim1} yields the integro-differential equation governing the two-dimensional horizontal motion~\citep{Oza2013}
\begin{widetext}
\begin{align}
m\ddot{\mathbf{x}}_d + D\dot{\mathbf{x}}_d =
\frac{mgAk_F}{T_F}
\int_{-\infty}^{t}
f\!\left(k_F|\mathbf{x}_d(t)-\mathbf{x}_d(s)|\right)
\frac{\mathbf{x}_d(t)-\mathbf{x}_d(s)}{|\mathbf{x}_d(t)-\mathbf{x}_d(s)|}
e^{-(t-s)/(T_F\mathrm{Me})}\,\mathrm{d}s,
\label{eq:dimensional2D}
\end{align}
\end{widetext}
where $f(\cdot)=-W'(\cdot)$ with prime denoting derivative with respect to its argument. Non-dimensionalizing using $\mathbf{x}'=k_F\mathbf{x}$ and $t'=Dt/m$ yields
\begin{align}\label{eq:dimless2D}
&\ddot{\mathbf{x}}_d + \dot{\mathbf{x}}_d = \\ \nonumber
&R\int_{-\infty}^{t}
f\!\left(|\mathbf{x}_d(t)-\mathbf{x}_d(s)|\right)
\frac{\mathbf{x}_d(t)-\mathbf{x}_d(s)}{|\mathbf{x}_d(t)-\mathbf{x}_d(s)|}
e^{-(t-s)/\tau}\,\mathrm{d}s,
\end{align}
with dimensionless wave amplitude $R=m^3gAk_F^2/(D^3T_F)$ and memory time $\tau=DT_F\mathrm{Me}/m$. 

For Fig.~\ref{fig:2D results}(a) in Appendix~\ref{sec:2D}, Eq.~\eqref{eq:dimless2D} is solved numerically using the wave form $W(|\mathbf{x}|)=\cos(|\mathbf{x}|)\exp[-(|\mathbf{x}|/L)^2]$ with $L=2\pi$, following the numerical scheme of Ref.~\citep{4cgg-hnyh} and time step $\Delta t=2^{-6}$. $10000$ different initial conditions were simulated for $t=100$ where the particles were initialized at random angular positions on a ring with radius $50$ dimensionless units that surrounds the Gaussian defect located at the origin, and the particles were directed, with small orientational noise, towards the center of the localized Gaussian defect with dimensionless width $W=1$ and dimensionless height $H=5$. For the circular corral configuration in Fig.~\ref{fig:2D results}(b), the same Gaussian potential barrier with $W=1$ and $H=5$ was extruded along a circle to create a ring barrier of radius $8\pi$. In this configuration, $10000$ particles were initiated uniformly inside the corral and directed radially outwards, with small noise in orientation, and evolved for a time of $t=200$.

\begin{figure*}[t]
\centering
\includegraphics[width=1.5\columnwidth]{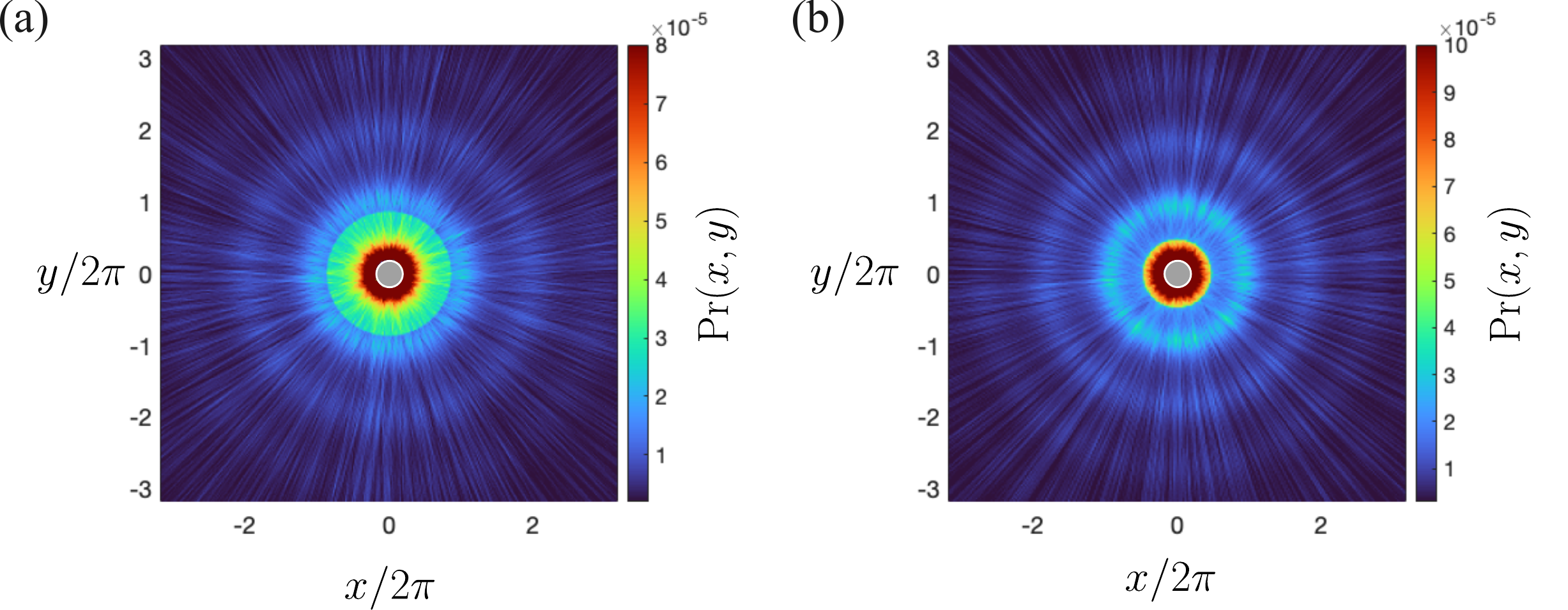}
\caption{Robustness of wave-like ensemble statistics to spatial wave form in walking droplet models. In $2$D stroboscopic model of walkers, particles are initialized far from a localized bump at the origin and directed towards the bump. This produces concentric Friedel-like oscillations in the probability density $\Pr(x,y)$ of outgoing trajectories for (a) Bessel wave for $W(|\mathbf{x}|)=J_0(|\mathbf{x}|)$ and $\tau=4$, and (b) $W(|\mathbf{x}|)=\cos(|\mathbf{x}|)$ and $\tau=2$. 
Color maps show $\Pr(x,y)$ constructed from $10^4$ outgoing trajectories. Parameters: $R=1$, $W=1$, $H=5$.}
\label{fig:robustness wave shape}
\end{figure*}

Restricting the dynamics to one spatial dimension yields
\begin{equation}
\ddot{x}_d + \dot{x}_d =
R\int_{-\infty}^{t}
f\!\left(x_d(t)-x_d(s)\right)
e^{-(t-s)/\tau}\,\mathrm{d}s,
\label{eq:dimless1D}
\end{equation}
which forms the basis of the minimal model studied in the main text. While experimentally measured wave fields are well approximated by Bessel functions~\citep{Oza2013,Molacek2013DropsTheory,Damiano2016}, the essential ingredient for capturing walking instabilities are spatial oscillations~\citep{ValaniUnsteady}. This feature is qualitatively captured by choosing $W(x)=\cos x$ and $f(x)=\sin x$, which enables an exact reduction of Eq.~\eqref{eq:dimless1D} to a low-dimensional Lorenz-like system~\citep{Durey2020lorenz,Valanilorenz2022}
\begin{equation}
\begin{aligned}
\dot{x}_d &= X,\\
\dot{X} &= Y - X,\\
\dot{Y} &= -\frac{Y}{\tau} + XZ,\\
\dot{Z} &= R - XY - \frac{Z}{\tau},
\end{aligned}
\label{eq:lorenz2}
\end{equation}
where
\begin{align}
Y(t)&=R\int_{-\infty}^{t}
\sin\!\left(x_d(t)-x_d(s)\right)
e^{-(t-s)/\tau}\,\mathrm{d}s,\\
Z(t)&=R\int_{-\infty}^{t}
\cos\!\left(x_d(t)-x_d(s)\right)
e^{-(t-s)/\tau}\,\mathrm{d}s.
\end{align}
Here, $X=\dot{x}_d$ is the particle velocity, $Y$ is the wave-memory force, and $Z$ is a dimensionless measure of the wave height at the particle location. Owing to our non-dimensionalization, the wavelength of the particle-generated waves is $2\pi$.

\section{Linear stability of the alternate internal-state model}
\label{sec:altstability}

We consider the alternate active-particle model introduced in Eq.~\eqref{eq:lorenz alt} with the external forcing removed ($H=0$),
\begin{align}
\dot{X}&=Y-X,\\
\dot{Y}&=-Y-XZ+\alpha X,\\
\dot{Z}&=\beta X^2-\gamma Z.
\end{align}

The steady-propulsion equilibria satisfy
\[
Y^*=X^*,\qquad
Z^*=\alpha-1,
\]
together with
\[
\beta (X^*)^2=\gamma(\alpha-1),
\]
yielding the pair of steadily propagating states
\[
(X^*,Y^*,Z^*)
=
\left(
\pm\sqrt{\frac{\gamma(\alpha-1)}{\beta}},
\pm\sqrt{\frac{\gamma(\alpha-1)}{\beta}},
\alpha-1
\right).
\]

Linearizing about either equilibrium gives the Jacobian
\[
J=
\begin{pmatrix}
-1 & 1 & 0\\
1 & -1 & -X^*\\
2\beta X^* & 0 & -\gamma
\end{pmatrix},
\]
whose characteristic polynomial is
\[
\lambda^3
+
(\gamma+2)\lambda^2
+
2\gamma\lambda
+
2\gamma(\alpha-1)
=0.
\]

For the parameters used in Fig.~\ref{fig:robustness internal-state},
\[
\alpha=4.1,\qquad
\beta=1,\qquad
\gamma=1.5,
\]
the eigenvalues are
\[
\lambda
=
-3.418,\qquad
-0.0409\pm1.649\,i,
\]
demonstrating that the steady-propulsion state is a stable spiral. The oscillation frequency of small perturbations is therefore
\[
\omega\simeq1.649.
\]

Assuming that the particle propagates with approximately the steady speed
\[
X^*
=
\sqrt{\frac{\gamma(\alpha-1)}{\beta}}
\simeq2.156,
\]
the corresponding spatial oscillation wavelength is estimated as
\[
\Lambda
\approx
\frac{2\pi X^*}{\omega}
\simeq8.21,
\]
or equivalently
\[
\frac{\Lambda}{2\pi}
\approx
1.31.
\]

This estimate is in good qualitative agreement with the wavelength of the oscillatory ensemble statistics observed numerically in Fig.~\ref{fig:robustness internal-state}. 

\section{Two-dimensional walking-droplet realization}
\label{sec:2D}

The active-particle mechanism developed in the main text is motivated by the reduced dynamics of walking droplets. To demonstrate that the observed wave-like ensemble statistics are not an artifact of the one-dimensional reduction, we consider the full two-dimensional stroboscopic pilot-wave model of walking droplets~\citep{Oza2013}. The particle dynamics are governed by the integro-differential trajectory equation described in Appendix~\ref{sec:1}, and ensemble statistics are computed using the spatially oscillatory and decaying wave form $W(|\mathbf{x}|)=\cos(|\mathbf{x}|)\exp[-(|\mathbf{x}|/L)^2]$.

Figure~\ref{fig:2D results} presents representative results in both open and confined geometries. In the open configuration (Fig.~\ref{fig:2D results}(a)), particles initialized far from a localized Gaussian bump produce concentric oscillatory modulations in the probability density $\Pr(x,y)$ reminiscent of Friedel-like statistics~\citep{Friedal}. In the confined geometry (Fig.~\ref{fig:2D results}(b)), particles initialized inside a circular corral generate structured wave-like ensemble patterns throughout the domain. Although these simulations are not intended as quantitative reproductions of the corresponding walking-droplet corral experiments~\citep{PhysRevE.88.011001,Saenz2017}, they demonstrate that the attractor-driven mechanism identified in the reduced model persists in the full two-dimensional stroboscopic dynamics.

Finally, we examine the robustness of these results to the detailed spatial form of the wave field. Figure~\ref{fig:robustness wave shape} compares ensemble statistics obtained using the experimentally motivated Bessel wave form and the simplified cosine wave form adopted in the main text. In both cases, concentric oscillatory probability densities emerge around the localized perturbation, indicating that the qualitative behavior is insensitive to the precise spatial profile of the underlying wave field.

\end{document}